\documentclass[,cup7b]{cupbook}

\usepackage{graphicx}

\begin{document}

%

\author[Rodolfo Gambini and Jorge Pullin]
{Rodolfo Gambini\\ 
Instituto de F\'{\i}sica, Facultad de Ciencias, Igu\'a 4225,
  Montevideo, Uruguay
\and
Jorge Pullin\\
Department of Physics and Astronomy, Louisiana State University\\
  Baton Rouge, LA 70803 USA}
\chapter{Consistent discretizations as a road to quantum gravity}

\begin{abstract}
  We present a brief description of the ``consistent discretization''
  approach to classical and quantum general relativity. We exhibit a
  classical simple example to illustrate the approach and summarize
  current classical and quantum applications. We also discuss the
  implications for the construction of a well defined quantum theory
  and in particular how to construct a quantum continuum limit.
\end{abstract}
  
\section{Consistent discretizations: the basic idea}

There has long been the hope that lattice methods could be used as a
non-perturbative approach to quantum gravity. This is in part based on
the fact that lattice methods have been quite successful in the treatment
of quantum chromodynamics.  However, one needs to recall that one of
the appeals of lattice methods in QCD is that they are gauge
invariant regularization methods. In the gravitational context this is
not the case. As soon as one discretizes space-time one breaks the
invariance under diffeomorphisms, the symmetry of most gravitational
theories of interest. As such, lattice methods in the gravitational
context face unique challenges. For instance, in the path integral
context, since the lattices break some of the symmetries of the
theory, this may complicate the use of the Fadeev--Popov technique.
In the canonical approach if one discretizes the constraints and
equations of motion, the resulting discrete equations are
inconsistent: they cannot be solved simultaneously. A related problem
is that the discretized constraints fail to close a constraint
algebra.

To address these problems we have proposed (Gambini \& Pullin 2003b,
Di Bartolo {\em et. al.} 2002) a different methodology for
discretizing gravitational theories (or to use a different terminology
``to put gravity on the lattice''). The methodology is related to a
discretization technique that has existed for a while in the context
of unconstrained theories called ``variational integrators'' (Lew {\em
  et. al. 2004}).  In a nutshell, the technique consists in
discretizing the action of the theory and working from it the discrete
equations of motion. Automatically, the latter are generically
guaranteed to be consistent. The resulting discrete theories have
unique features that distinguish them from the continuum theories,
although a satisfactory canonical formulation can be found for them
(Di Bartolo {\em et. al.} 2005). The discrete theories do not have
constraints associated with the space-time diffeomorphisms and as a
consequence the quantities that in the continuum are the associated
Lagrange multipliers (the lapse and the shift) become regular
variables of the discrete theories whose values are determined by the
equations of motion. We call this approach in the context of
constrained theories ``consistent discretizations''.

The consistently discretized theories are both puzzling and
attractive. On the one hand, it is puzzling that the Lagrange
multipliers get fixed by the theory. Don't the Lagrange multipliers
represent the gauge freedom of general relativity? The answer is what
is expected: the discretization breaks the freedom and solutions to
the discrete theory that are different correspond, in the continuum
limit, to the same solution of the continuum theory. Hence the
discrete theory has more degrees of freedom.  On the other hand, the
lack of constraints make the consistently discretized theories
extremely promising at the time of quantization. Most of the hard
conceptual questions of quantum gravity are related to the presence of
constraints in the theory.  In comparison, the consistently
discretized theories are free of these conceptual problems and can be
straightforwardly quantized (to make matters even simpler, as all
discrete theories, they have a finite number of degrees of freedom).
In addition, they provide a framework to connect the path integral and
canonical approaches to quantum gravity since the central element is a
unitary evolution operator. In particular they may help reconcile the
spin foam and canonical loop representation approaches. They also
provide a natural canonical formulation for Regge calculus 
(Gambini \& Pullin 2005c).

In this article we would like to briefly review the status of the
consistent discretization approach, both in its application as a
classical approximation to gravitational theories and as a tool for
their quantization.  Other brief reviews with different emphasis can
be seen in Gambini \& Pullin (2005a, 2005b). The organization of the
article is as follows. In section \ref{section2} we consider the
application of the technique to a simple, yet conceptually challenging
mechanical model and discuss how features that one observes in the
model are actually present in more realistic situations involving
general relativity. In section \ref{section3} we outline various
applications of the framework.  In section \ref{section4} we discuss
in detail the quantization of the discrete theories and in section
\ref{section5} we outline how one can define the quantum continuum
limit.  We end with a summary and outlook.

\section{Consistent discretizations}
\label{section2}

To introduce an illustrate the method in a simple ---yet
challenging--- model we consider the model analyzed in detail by
Rovelli (1990) in the context of the problem of time in
canonical quantum gravity: two harmonic oscillators with constant
energy sum. We have already discussed this model in some detail
in Gambini \& Pullin (2005b) but we would like to revisit it here
to frame the discussion with a different emphasis.

The model has canonical coordinates $q^1,q^2,p^1,p^2$ with
the standard Poisson brackets and a constraint given by,
\begin{equation}
  C=\frac{1}{2} \left((p^1)^2+(p^2)^2+(q^1)^2+(q^2)^2\right)-M=0,
\end{equation}
with $M$ a constant. The model is challenging since no standard
unconstrained Hamiltonian formulation can correspond to this dynamical
system since the presymplectic space is compact and therefore cannot
contain any $S\times R$ structure. Nevertheless, we will see that the
consistent discretization approach does yield sensible results. This
helps dispel certain myths about the consistent discretization
scheme. Since it determines Lagrange multipliers, a lot of people tend
to associate the scheme with some sort of ``gauge fixing''.  For this
model however, a gauge fixing solution would be unsatisfactory, since
it would only cover a portion of phase space. We will see that this is not
the case in the consistent discretization scheme. We will also see
that the evolution scheme is useful numerically in practice.

We start by writing a discrete Lagrangian for the model,
\begin{eqnarray}
L(n,n+1) &=& 
p^1_n \left( q^1_{n+1}-q^1_n\right)+
p^2_n \left( q^2_{n+1}-q^2_n\right)\\
&&-\frac{N_n}{2}\left(
(p^1_n)^2+(p^2_n)^2+(q^1_n)^2+(q^2_n)^2-2 M\right),\nonumber
\end{eqnarray}
and working out the canonical momenta for all the variables, i.e.,
$P_q^1$, $P_q^2$, $P_p^1$, $P_p^2$. The momenta of a variable at level $n$ are
obtained by differentiating $L(n,n+1)$ with respect to the variable
at level $n+1$. One then eliminates the $p^{1,2}$ and
the $P_p^{1,2}$ and is left with evolution equations for the canonical
pairs,
\begin{eqnarray}
  q^1_{n+1} &=& q^1_n+N_n \left(P^1_{q,n}-2 q^1_n\right)\\
  q^2_{n+1} &=& q^2_n+N_n \left(P^2_{q,n}-2 q^2_n\right)\\
  P^1_{q,n+1} &=&  P^1_{q,n} -N_n q^1_n\\
  P^2_{q,n+1} &=&  P^2_{q,n} -N_n q^2_n.
\end{eqnarray}

The Lagrange multiplier gets determined by the solution(s) of a 
quadratic equation that is obtained working out the momenta of the
Lagrange multipliers, 
\begin{eqnarray}
&&\left((q^1_n)^2+(q^2_n)^2\right)(N_n)^2-2\left(P^1_{q,n} q^1_n+
P^2_{q,n} q^2_n\right)N_n+\nonumber\\ 
&&+\left(  P^1_{q,n}\right)^2+\left(  P^2_{q,n}\right)^2
+\left(q^1_n\right)^2+\left(q^2_n\right)^2-2M=0.
\label{quadratic}
\end{eqnarray}

The resulting evolution scheme when one eliminates the Lagrange
multipliers using equation (\ref{quadratic}) constitutes a canonical
transformation between instants $n$ and $n+1$. This result may appear
puzzling at first, a general discussion of how this can be framed in a
Dirac-like approach for discrete theories can be seen in
Di Bartolo {\em et. al.} (2005).

We would like to use this evolution scheme to follow numerically the
trajectory of the system. For this, we need to give initial data.
Notice that if one gives initial data that satisfy the constraint
identically at level $n$, the quadratic equation for the lapse has a
vanishing independent term and therefore the solution is that the
lapse $N$ vanishes (the non-vanishing root will be large and would
imply a large time evolution step that puts us away from the continuum
generically).  To construct initial data one therefore considers a set
for which the constraint vanishes and introduces a small perturbation
on one (or more) of the variables. Then one will have evolution.
Notice that one can make the perturbation as small as desired. The
smaller the perturbation, the smaller the lapse and the closer the
solution will be to the continuum.

For concreteness, we choose the following initial values for the
variables, $M=2, q^1_0 =0, q^2_0 =(\sqrt{3}-\Delta) \sin({\pi\over
4}), P^1_{q,0} = 1, P^1_{q,0} = (\sqrt{3}-\Delta) \cos({\pi\over 4})$.

We choose the parameter $\Delta$ to be the perturbation, i.e.,
$\Delta=0$ corresponds to an exact solution of the constraint, for
which the observable $A=1/2$ (see below for its definition). The
evolution scheme can easily be implemented using a computer algebra
program like Maple or Mathematica.

Before we show results of the evolution, we need to discuss in some
detail how the method determines the lapse. As we mentioned it is
obtained by solving the quadratic equation (\ref{quadratic}). This
implies that for this model there will be two possible
solutions and in some situations they could be negative or
complex. One can choose any of the two solutions at each point during
the evolution. This ambiguity can be seen as a remnant of the
re-parameterization invariance of the continuum. It is natural
numerically to choose one ``branch'' of the solution and keep with
it. However, if one encounters that the roots become complex, we have
observed that it is possible to backtrack to the previous point in the
iteration, choose the alternate root to the one that had been used up
to that point and continue with the evolution. A similar procedure could
be followed when the lapse becomes negative. It should be noted that
negative lapses are not a problem per se, it is just that the
evolution will be retraced backwards. We have not attempted to correct
such retracings, i.e. in the evolutions shown we have only ``switched
branches'' whenever the lapse becomes complex. This occurs when the
discriminant in the quadratic equation (\ref{quadratic}) changes sign.

We would like to argue that in some sense the discrete model
``approximates'' the continuum model well. This, however, turns out to
be a challenging proposition in re-parameterization invariant
theories. The first thing to try, to study the evolution of the
quantities as a function of $n$ is of course meaningless as a grounds
to compare with the continuum. In the discrete theory we do not
control the lapse, therefore plots of quantities as a function of $n$
are meaningless. To try to get more meaningful information one would
like to concentrate on ``observables''.  In the continuum theory,
these are quantities that have vanishing Poisson brackets with the
constraints (also sometimes known as ``perennials''). Knowing these
quantities as functions of phase space allows to know any type of
dynamical physical behavior of the system. One can use them, for
instance, to construct ``evolving constants'' (Rovelli 1990). The
existence of perennials in the continuum theory is associated with
symmetries of the theory.  If such symmetries are not broken by the
discretization process, then in the discrete theory one will have
exact conserved quantities that correspond to the perennials of the
continuum theory. The conserved quantities will be given by
discretizations of the perennials of the continuum. It should be noted
that in the continuum theory perennials as functions of phase space
are defined up to the addition of multiples of the constraints. There
are therefore infinitely many versions of a given perennial. When
discretized these versions are inequivalent (since in the discrete
theory the constraints of the continuum theory do not hold exactly)
and only one of these versions will correspond to an exact conserved
quantity of the discrete theory.

In this model there are two independent perennials in the 
continuum. One of them becomes straightforwardly upon discretization
an exact conserved quantity of the discrete theory,
\begin{equation}
  O_1=p^1 q^2-p^2 q^1.
\end{equation}

\begin{figure}[htbp]
  \centerline{\includegraphics[height=6.2cm]{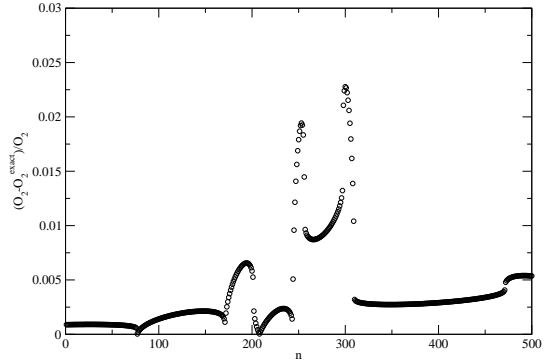}}
    \caption{The model has two ``perennials". One of them is an exact
    conserved quantity of the discrete theory, so we do not present a
    plot for it. The second perennial $(O_2)$ is approximately
    conserved. The figure shows the relative error in its computation
    in the discrete theory. It is worthwhile noticing that, unlike
what is usual in free evolution schemes, errors do not accumulate,
    they may grow for a while but later they might diminish. }
  \label{figure4}
\end{figure}

Another perennial is given by
\begin{equation}
  O_2=(p^1)^2-(p^2)^2+(q^1)^2-(q^2)^2.
\end{equation}
This quantity is not an exact conserved quantity of the discrete
model, it is conserved approximately, as we can see in figure
(\ref{figure4}).  We at present do not know how to find an exact
conserved quantity in the discrete theory that corresponds to a
discretization of this perennial (plus terms proportional to the
constraint). In the end, this will be the generic situation, since in
more complicated models one will not know exact expressions either for
the perennials of the continuum theory or the constants of motion of
the discrete theory. Notice also that in the continuum, in order to
recover physical information about the system, one generically needs
the two perennials plus combinations involving the constraints. In the
discrete theory these combinations will not be exactly
preserved. Therefore even if we found exact conserved quantities for
both perennials in the discrete theory, the extracted physics would
still only be approximate, and the measure of the error will given by  how
well the constraint of the continuum theory is satisfied in the
discrete theory. It is in this sense that one can best say that the
discrete theory ``approximates the continuum theory well''.

Figure (\ref{figure4}) depicts the relative errors throughout
evolution in the value of the second perennial we discussed.
Interestingly, 
although in intermediate steps of the evolution the error
grows, it decreases later.

As we argued above, in the discrete theory quantities approximate the
ones of the continuum with an error that is proportional to the value
of the constraint. Therefore the value of the constraint is the real
indicator of how accurately one is mirroring the continuum theory. It
is a nice feature to have such an error indicator that is independent
of the knowledge of the exact solution. Using this indicator one can,
for instance, carry out convergence studies and show that the method
does indeed converge for this model in a detailed way
(Gambini \& Pullin 2005b).

\begin{figure}[htbp]
  \centerline{\includegraphics[height=8.2cm]{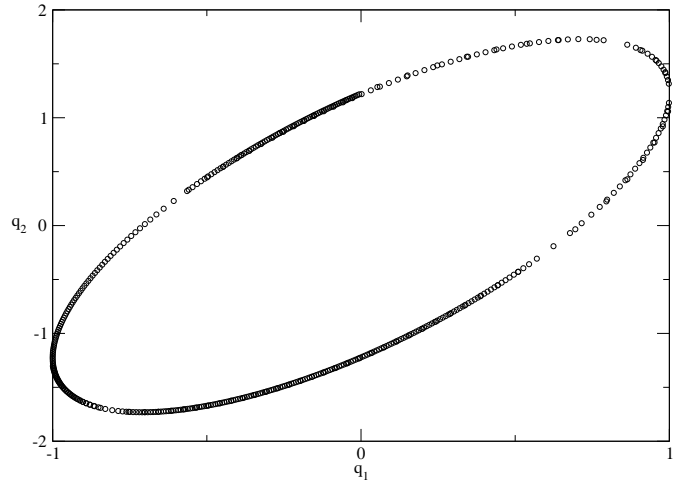}} 
\caption{The orbit
    in configuration space. As it is readily seen, the consistent
    discrete approach covers the entire available configuration space.
    This clearly exhibits that the approach is not a ``gauge fixing''.
    Gauge fixed approaches cannot cover the entire configuration space
    due to its compact nature. The dynamical changes in the value of
    the lapse can be seen implicitly through the density of points in
    the various regions of the trajectory. Also apparent is that the
    trajectory is traced on more than one occasion in various
    regions. Deviation from the continuum trajectory is not noticeable
    in the scales of the plot.}
  \label{figure3} \end{figure}

Figure (\ref{figure3}) shows the trajectory in configuration space.
As we see, the complete trajectory is covered by the discretized
approach.  This is important since many people tend to perceive the
consistent discretization approach as ``some sort of gauge fixing''.
This belief stems from the fact that when one gauge fixes a theory,
the multipliers get determined. In spite of this superficial analogy,
there are many things that are different from a gauge fixing. For
instance, as we discussed before, the number of degrees of freedom
changes (for more details see Gambini \& Pullin 2003c).  In addition
to this, this example demonstrates another difference. If one indeed
had gauge fixed this model, one would fail to cover the entire
available configuration space, given its compact nature.

To conclude this section, let us point out to some hints
that this model provides. To begin with, we see that the consistent
discretization scheme successfully follows the classical 
continuum trajectory. One has control of how accurate things 
are by choosing the initial data. One can show that the approach
converges using estimators of error that are independent of knowledge
of exact solutions or other features generically not available. 
The solution of the equations for the Lagrange multipliers may
develop branches, and one can use this to one's advantage in 
tackling problems where the topology of phase space is not 
simple.

What is the state of the art in terms of applying this approach as a
classical numerical relativity tool? We have applied the method in
homogeneous cosmologies and also in Gowdy cosmologies (Gambini, Ponce
\& Pullin 2005) where one has spatial dependence of the variables. All
of the features we have seen in the model described in this section
are present in the more complicated models, the only difference is
computational complexity. How well does it compete with more
traditional numerical relativity approaches? At the moment the method
is too costly to compete well, since the evolution equations are
implicit. But as traditional ``free evolution'' methods in numerical
relativity keep on encountering problems of instabilities and
constraint violations, and as computational power increases, the
costliness of the consistent discretization approach may become less
of a problem. A challenge to be overcome is that in situations of
interest the problems have boundaries, and the approach has not yet
been worked out in the presence of boundaries, although we are
actively considering this point.

\section{Applications}
\label{section3}
\subsection{Classical relativity}

As we argued before, our approach can be used to construct discrete
theories that approximate general relativity. It is therefore suitable
for doing numerical relativity. The main problem is that the resulting
numerical schemes are implicit, and therefore very costly in
situations of physical interest where there are no symmetries. Most of
present numerical relativity is being pursued with explicit algorithms
for that reason.  In spite of this, our experience with the model
analyzed by Rovelli and the Gowdy cosmologies indicates that our
discretizations may have attractive features that are not present in
more traditional discretization schemes. In particular the fact that
errors do not seem to accumulate but rather grow and decrease in
cycles as one evolves, could offer unique promises for long term
evolutions like the ones desired in binary systems that emit
gravitational waves.  In addition to this, it has been shown (Di
Bartolo {\em et. al.} 2005a) that our approach applied to linearized
gravity yields a discretization that is ``mimetic'', that is, the
constraints are automatically preserved without determining the
Lagrange multipliers. This may suggest that at least at linearized
level, our discretizations may perform better than others.

In spite of these hints of a promise, there is a lot of terrain yet to
cover before one could consider seriously using one of these schemes
in problems of current interest. In particular, it has growingly been
recognized in numerical relativity the importance of having symmetric
hyperbolic formulations (see Reula (1998) for a review) and in
particular of incorporating constraint preserving boundary
conditions. Most symmetric hyperbolic formulations are constructed at
the level of equations of motion and do not derive from an action
principle. Therefore our discretization technique is not directly
applicable. More work is clearly needed in this area.

Another area of recent progress (Gambini \& Pullin 2005c) has been the
application of these ideas to Regge calculus. In Regge calculus it had
been observed that the canonical formulation was problematic. In
particular it seemed to require that the Lagrange multipliers be fixed
(Friedman \& Jack 1986). This is exactly the statement that we use as
a starting point for our discrete construction. We have recently shown
how one can construct an unconstrained version of canonical Regge
calculus in which some of the lengths of the links are determined
precisely mirroring what happens with the Lagrange multipliers in
other theories. Although this is only a beginning, it suggests a novel
technique to have a canonical formulation of Regge calculus that may
have attractive implications quantum mechanically (for instance it
contains a new prescription to define the path integral).

\subsection{The problem of time}

Since the discrete theory that one constructs through our procedure is
constraint-free, it immediately circumvents most of the hard
conceptual problems of canonical quantum gravity including the
``problem of time''. The issue is a bit more subtle than it initially
appears. One indeed has a theory without constraints and a ``genuine
evolution'', except that the latter is cast in terms of the discrete
parameter $n$. This parameter cannot be accessed physically, it is not
one of the variables one physically observes for the systems under
study. This forces us to consider a ``relational'' formulation, in the
same spirit as Page and Wootters (1983) considered . The idea is
to pick one of the physical variables and use it as a clock. One then
asks relational questions, for instance ``what is the conditional
probability than one of the other variables takes a given value when
the clock variable indicates a certain time''. These questions can of
course also be asked in continuum general relativity, but the detailed
construction of the conditional probabilities is problematic, due to
the difficulties of having a probabilistic interpretation of quantum
states in canonical quantum gravity (see the discussion in
Kucha\v{r} 1992). In our approach, on the other hand, the conditional
probabilities are well defined, since there are no constraints to
generate problems with the probabilistic interpretation of states.
For more details see (Gambini, Porto \& Pullin 2003).

\subsection{Cosmological applications}

We have applied the technique to cosmological models. The use of these
discrete theories in cosmology has an attractive consequence. Since
the lapse, and therefore the ``lattice spacing in time'' is determined
by the equations of motion, generically one will avoid the singularity
classically. Or to put it in a different way, one would have to ``fine
tune'' the initial data to reach the singularity (unless one uses
variables in which the singularity is on a boundary of phase
space). Quantum mechanically, this implies that the singularity will
be probabilistically suppressed. As the discrete theory 
tunnels through the singularity, there is a precise sense in which
one can claim that  the lattice spacing changes qualitatively. This
could be used to argue that physical constants change when tunneling
through a singularity since in lattice theories the ``dressed'' value
of the coupling constants is related to the lattice spacing. Therefore
this provides a concrete mechanism for Smolin's ``The life of the cosmos''
proposal (1992). For more details see Gambini \& Pullin (2003a).

\subsection{Fundamental decoherence, black hole information puzzle, limitations to quantum computing}

Once one has solved the problem of time in the relational fashion
discussed above, one notices that the resulting quantum theory
fails to be unitary. This is reasonable. In our approach, when one
quantizes, one would have a unitary evolution of the states as a 
function of the discrete parameter $n$. In the relational approach
one picked some dynamical variable and called it time $T$. 
Suppose one chose a state in which this variable is highly peaked as
a function of $n$. If one lets the system evolve, the variable will
spread and at a later instant one would have a distribution of values
of $n$ that correspond to a given $T$ (or vice-versa). That means
that if one started with a ``pure'' state, one ends with a mixed
state. The underlying reason is that the physical clock $T$ cannot
remain in perfect lock-step with the evolution parameter $n$.

A detailed discussion of the implications of this lack of unitarity is
in Gambini, Porto \& Pullin (2004a, 2004b, 2005a). Of course, this is
not the first time that quantum gravity effects have been associated
with loss of unitarity.  However, unlike previous proposals (see Banks
{\em et. al} 1984), the detailed evolution implied by the relational
description we find conserves energy, which is a very desirable
feature.  One can give a bound on the smallness of the effect by
taking into account what is the ``best'' clock one can construct from
fundamental physical principles (Ng \& van Dam 1995). The lack of
unitarity makes the off diagonal elements of the density matrix go to
zero exponentially. The exponent (for a system with two energy levels,
for simplicity) is proportional to minus the Bohr frequency between
the levels squared, to the Planck time to the $(4/3)$ power and to the
time one waits for the state to lose coherence to the $(2/3)$ power
(these results appear not even to be Galilean invariant, but this is
not the case as discussed in detail in Gambini, Porto \& Pullin
2004c). It is clear that the effect is negligible for most quantum
systems. Chances of observing the effect in the lab (see for instance
Simon \& Jaksch 2004) are at the moment remote, one would require a
quantum system of macroscopic size. If one assumes energy differences
of $eV$ size, one would roughly need $10^{13}$ atoms. Bose-Einstein
condensates at present can achieve states of this sort with perhaps
hundreds of millions of atoms, but they do not involve energy
differences of $eV$'s per atom.  Another important caveat of these
types of discussions is that they have been carried out at a very
naive level of Newtonian quantum mechanics. If one were to consider
relativistic quantum field theory, one would have to have a ``clock''
variable per spatial point. This would imply that quantum states would
lose coherence not only as time evolves, but also between points in
space.  Such effects could potentially have consequences that are much
more amenable to experimental testing (Simon \& Jaksch 2004).

Once one accepts that quantum mechanics at a fundamental level
contains loss of unitarity one may wish to reconsider the black hole
information paradox. After all, the reason one has a paradox is that
when a black hole evaporates, the final result is a mixed state, even
if one built the black hole by collapsing a pure state.  The question
is: does this loss of unitarity occur faster or slower than the one we
have found? If it is slower, then it will be unobservable.  A priori
one could expect that the effect we discussed should not be too
important. We just argued in the previous paragraph that it is very
small. However, black holes take a long time to evaporate.  And as
they evaporate their energy levels become more separated as the
temperature increases. A detailed calculation shows that the order of
magnitude of the off-diagonal elements of the density matrix at the
time of complete evaporation would be approximately $M_{BH}^{-2/3}$
with $M_{BH}$ the black hole mass in Planck mass units (Gambini, Porto
\& Pullin 2005a). For an astrophysical size black hole therefore the
loss of unitarity is virtually complete and the paradox cannot be
realized physically. What happens if one takes, say, a very small
black hole? Can one reformulate the paradox in that case? The
formulation we have is not precise enough to answer this question. We
have only roughly estimated the magnitude of the decoherence just to
give an order of magnitude estimate. Many aspects of the calculation
are also questionable for small black holes, where true quantum
gravity effects are also important.

An interesting additional observation (Gambini, Porto \& Pullin 2005b)
is that the loss of quantum coherence we found can provide a
fundamental limitation to how fast quantum computers can operate that
is more stringent than other fundamental limits considered.

\section{Constructing the quantum theory}
\label{section4}

As we argued above, the construction of the quantum theory starts by
implementing the canonical transformation that gives the evolution in
terms of the discrete parameter $n$ as a unitary transformation.
Before doing this one constructs the canonical theory that results
from the elimination of the Lagrange multipliers. The resulting
canonical theory generically has no constraints, and has evolution
equations for its canonical variables. One picks a polarization, for
instance $\Psi(q)$ where $q$ is a set of configuration variables, and
considers the unitary transformation as operating on the space of
wavefunctions chosen. Since generically there are no constraints, one
can pick as physical inner product the kinematical one and construct a
Hilbert space of wavefunctions that are square integrable.  
If one is in the Schr\"odinger representation states
evolve, so we label them as $\Psi_n(q)$ and the evolution is given by,
\begin{equation}
\Psi_{n+1}(q)=\int dq' U(q|q') \Psi_n(q').
\end{equation}  
The transformation has to be such that it implements the evolution
equations as operatorial relations acting on the space of
wavefunctions in the Heisenberg representation, where
\begin{equation}
U(q|q')=<n+1,q'|n,q>,
\end{equation}
and where $|n+1,q'>$ and $|n,q>$ are the eigenvectors of the 
configuration operators $\hat{q}$ in the Heisenberg representation
at levels $n+1$ and $n$ respectively. The evolution equations take
the form,
\begin{eqnarray}
<n+1,q| \hat{q}_{n+1} -f(\hat{q}_n,\hat{p}_n) |n,q'>&=&0,\\
<n+1,q| \hat{p}_{n+1} -g(\hat{q}_n,\hat{p}_n) |n,q'>&=&0,
\end{eqnarray}
with $f$, $g$ the quantum evolution equations, which are chosen to 
be self-adjoint in order for the transformation to be unitary.
Explicit examples of this construction for cosmological models can
be seen in (Gambini \& Pullin 2003c).

If at the end of this process one has constructed a transformation that
is truly unitary the quantization is complete in the discrete space and
one has a well defined framework to rigorously compute the conditional
probabilities that arise when one uses a relational time to describe
the physical system. This is a major advantage over attempts to construct
the relational picture with systems where one has constraints.

There are some caveats to this construction that are worth pointing
out.  As we mentioned, our construction generically yields discrete
theories that are constraint-free. To be more precise, the theories do
not have the constraints associated with space-time diffeomorphisms.
If the theory under consideration has other symmetries (for instance
the Gauss law of Yang--Mills theory or gravity written in the new
variable formulation), such symmetries may be preserved upon
discretization (we worked this out explicitly for Yang--Mills and BF
theory in Di Bartolo {\em et. al} 2002). The resulting discrete theory
therefore will have some constraints. If this is the case, the above
construction starts by considering as wavefunctions states that are
gauge invariant and endowed with a Hilbert space structure given by a
gauge invariant inner product. The resulting theory has true (free)
Lagrange multipliers associated with the remaining constraints. The
unitary transformation will depend on such parameters. An alternative
is to work in a representation where the constraints are solved
automatically (like the loop representation for the Gauss law). There
one has no constraints left and the inner product is the kinematical
one in the loop representation and the unitary transformation does not
depend on free parameters.

Other issues that may arise have to do with the fact that in many
situations canonical transformation do not correspond quantum
mechanically to unitary transformations. This problem has been
discussed, for instance, by Anderson (1994). He noted that
the only canonical transformations that can be implemented as unitary
transformations are those that correspond to an isomorphism of a phase
space into itself. This is important for the discrete theories in the
following way. If one has a continuum constrained theory, its physical
phase space is on the constraint surface. The discrete theories have a
phase space that includes the constraint surface of the continuum
theory. However, the discrete phase space variables cover only a
subspace of the kinematical phase space of the continuum theory.
There are inaccessible sectors that correspond to complex values of
the Lagrange multipliers in the discrete theory. Therefore, in order
to have the canonical transformation of the discrete theory be an
isomorphism, one may have to choose a physical Hilbert space for the
discrete theory that is a subspace of the kinematical space instead of
just taking it to be coincident.  This has to be done carefully, since
restricting the Hilbert space may imply that some physical quantities
fail to be well defined in the physical Hilbert space.  We have
explored some of these issues in some quantum mechanical models that
have a relational description. We have shown that one can successfully
recover the traditional quantum mechanical results in a suitable
continuum limit by carefully imposing a restriction on the kinematical
Hilbert space, and that one can define variables that approximate any
dynamical variable of the continuum theory in the continuum limit in
the restricted Hilbert space (see Di Bartolo {\em et. al} 2005b).

\section{The quantum continuum limit}
\label{section5}

As we argued in the discussion of the model analyzed by Rovelli, a
good measure of how close one is to the continuum theory in a given
solution of the discrete theory is to evaluate the constraint of the
continuum theory. Such constraint is only exactly satisfied in the
continuum limit. An alternative way of presenting this is to consider
the construction of a ``Hamiltonian'' such that exponentiated would
yield the unitary evolution between $n$ and $n+1$,
$\hat{U}=\exp(i\hat{H})$ where $\hbar=1$ and $\hat{H}$ has units of
action. Such Hamiltonian can only be constructed locally since in some
points of the evolution the logarithm of the unitary transformation is
not well defined. Such Hamiltonian can be written as a formal
expansion in terms of the constraint of the continuum theory (a way of
seeing this is to notice that in the continuum limit this Hamiltonian
has to vanish since it incorporates the timestep). If one chooses an
initial state such that $<\hat{H}>\ll 1$ the evolution will preserve
this ($\hat{H}$ is an exact constant of the motion). This will
continue until one reaches a point where $\hat{H}$ is not well
defined.  The evolution will continue, but it will not necessarily
remain close to the continuum limit. In certain cosmological examples
this point coincides with the point where the continuum theory has the
singularity, for example (Gambini \& Pullin 2003c). Therefore a first
condition on the quantum states in the continuum limit $<\hat{H}>\ll
1$. A second condition is that the expectation values of the physical
variables should not take values in the points where $\hat{H}$ is not
well defined. A third condition is not to make measurements with ``too
much accuracy'' on variables that do not commute with $\hat{H}$ . This
requirement stems from the fact that such measurements would introduce
too much dispersion in $\hat{H}$ and one would violate the first
requirement.  In examples we have seen that this condition translates
in not measuring $q,p$ with sharper accuracy than that of the step of
the evolution in the respective variable. This appears reasonable, a
discrete theory should not allow the measurement of quantities with
accuracies smaller than the discretization step.  The variables that
do not commute with $\hat{H}$ play a crucial role in the relational
description since they are the variables that can be used as
``clocks'' as they are not preserved under evolution as constants of
the motion.

\section{Summary and outlook}
\label{section6}

One can construct discrete canonical theories that are constraint free
and nevertheless approximate continuum constrained theories in a well
defined sense. The framework has been tested at a classical level in a
variety of models, including gravitational ones with infinitely many
degrees of freedom. Further work is needed to make the framework
computationally competitive in numerical relativity. In particular the
use of better discretizations in time, including higher order ones,
appears as promising. Initial explorations we are carrying out in
simple models indicate that one can achieve long-term stable and
accurate evolutions using moderately large timesteps. This could be
very attractive for numerical relativity if it turns out to be a
generic property.

Since the discrete theories are constraint free, they can be quantized
without serious conceptual obstacles. In particular a relational time
can be introduced in a well defined way and quantum states exhibit a
non-unitary evolution that may have implications experimentally and
conceptually (as in the black hole information puzzle). There is a 
reasonable proposal to construct the quantum continuum limit that has
been tested in simple constrained models. The main challenge is 
to apply the framework at a quantum level in systems with field theoretic
degrees of freedom. The fact that one has a well defined framework
that is computationally intensive suggests that this is an avenue
for conducting numerical quantum gravity.

This work was supported in part by grants NSF-PHY0244335,
NASA-NAG5-13430 and funds from the Horace Hearne Jr. Laboratory for
Theoretical Physics and CCT-LSU.


\end{document}